\documentclass[prc,aps,preprint,showkeys,showpacs,nofootinbib]{revtex4}
\usepackage{graphicx}

\begin{document}

\date{\today}

\title{\textbf{Production of complex particles in low energy spallation 
and in fragmentation reactions by in-medium random clusterization.}}
\author{Denis Lacroix and Dominique Durand}

\address{Laboratoire de Physique Corpusculaire, \\
ENSICAEN and Universit\'e de Caen,IN2P3-CNRS,\\
Blvd du Mar\'{e}chal Juin
14050 Caen, France}

\begin{abstract}

Rules for in-medium complex particle production 
in nuclear reactions are proposed. These rules have been implemented in 
two models to simulate nucleon-nucleus 
and nucleus-nucleus reactions around the Fermi energy \cite{Lac04,Lac05}.
Our work emphasizes the effect of randomness in cluster formation, 
the importance 
of the nucleonic Fermi motion as well as the role of conservation laws.
The concepts of total available phase-space 
and explored phase-space under constraint imposed by the reaction are clarified.        
The compatibility of experimental observations 
with a random clusterization 
is illustrated in a schematic scenario of a proton-nucleus collision. The 
role of randomness under constraint is also illustrated
in the nucleus-nucleus case.
\end{abstract}

\maketitle

\section{Introduction}

Nuclear reactions around the Fermi energy have revealed that nuclei
can break into 
several pieces of various sizes: the so-called multifragmentation process\cite{Dur01}. A 
striking feature of experimental 
observation is the large number of charge and energy partitions that can 
be accessed. In order to 
understand the statistical aspects of the explored phase-space, several 
physical origins have been proposed. 
Among them, the nuclear liquid-gas phase transition appears as one of
the best candidate. 
However, due to the complexity of nuclear reactions including impact 
parameter mixing, pre-equilibrium emission 
and thermal decay, it is hard to trace-back the process of 
cluster formation.
Nowadays, the complexity of experimental analyses increases 
constantly \cite{Iwm03, Ada04}. Conjointly, more and more
elaborated models have been proposed to simulate reactions \cite{Ayi88,Ran90,Gua96,Bon95,Gro90,Dan91,
Sto86,Aic91,Fel89,Ono98}. However, 
the issue concerning cluster formation remains a highly debated 
question. In this
work, we would like to contribute 
to the discussion on particle emission during the pre-equilibrium stage.
We have tested a large number of hypothesis 
for the formation of clusters
in the nuclear medium in order to provide event generators for the study of 
nuclear reactions.  
Guided by the experimental observation, surprising 
conclusions concerning 
the way cluster are formed may be assessed. Simple rules have
been found for the formation and the emission of complex particles. The 
hypothesis retained are not only fully compatible with 
experiments on multifragmentation, 
but seems also to be adequate for nucleon-nucleus reactions 
in the same energy range. 

The paper is organized as follows: first the rules for cluster formation and 
emission are introduced and illustrated in a schematic scenario for experiments. 
In a second part, additional effects that should be accounted to compare 
quantitatively with measurements are discussed. Finally, the compatibility
of the rules with data are illustrated. Conclusions and perspectives are drawn at the end of the
paper.

\section{Rules for the formation and the emission of clusters}

\begin{figure}[tbph]
%\begin{center}
\includegraphics[height=15.cm,angle=-90.]{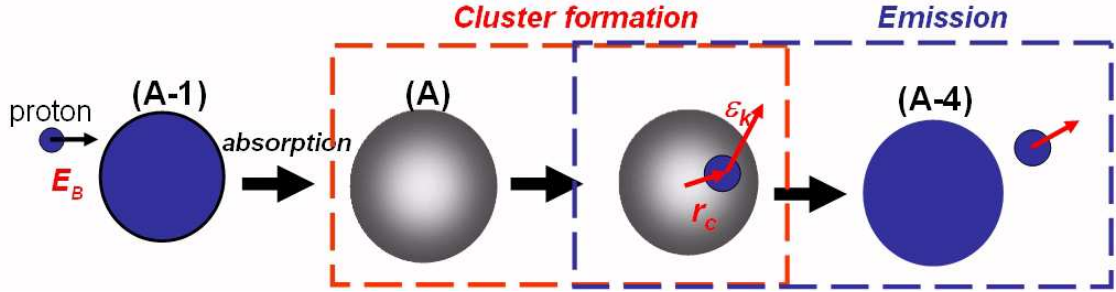}
%\end{center}
\caption{\it Schematic representation of a three step nucleon-induced 
reaction. A nucleon with beam energy close to the Fermi energy is 
first absorbed by a nucleus. Then two steps are identified for pre-equilibrium emission: the 
formation of the cluster and its emission.}
\label{fig:clus1}
\end{figure}

In this section, rules for the cluster formation and the production of
fragmentation 
partitions are defined. 
In order to illustrate these rules we consider a proton colliding a heavy 
target with an energy close to the Fermi energy
\footnote{The energy is chosen small enough to avoid strong 
influence of direct 
two-body nucleon-nucleon collisions}. Then a simplified three steps 
scenario is considered (see figure \ref{fig:clus1}). First the incident 
nucleon is absorbed. The second 
step corresponds to the in-medium formation of the cluster 
while the last step is the 
emission in the continuum.

\begin{figure}[tbph]
\includegraphics[height=7.cm]{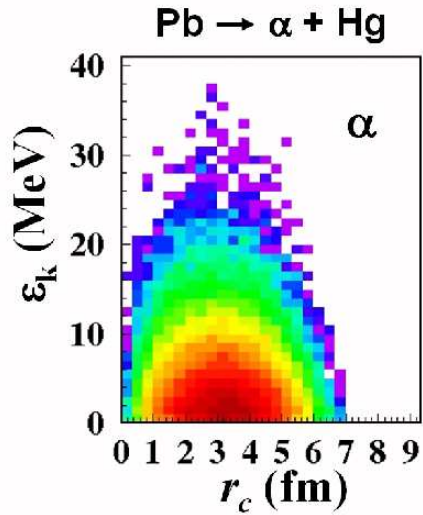}
\caption{ Correlation between the position and the 
kinetic energy per nucleon
for the $\alpha$ particles using a random sampling assumption for the 
nucleons forming the $\alpha$ particle. This two-dimensional plot 
corresponds 
to the total "accessible" phase-space for the considered particle.}
\label{fig:clus0}
\end{figure}

The hypothesis retained to describe the last two stages 
can be summarized as follow:
\begin{itemize}
\item {\bf Cluster formation:} Considering a cluster of mass $A_c$ and 
charge $Z_c$ formed in the medium, 
we assume that the cluster is composed of nucleons chosen randomly 
in the target\footnote{In the energy range considered here, a 
Thomas-Fermi distribution corresponding to the ground state of the target is
assumed. This means that in-medium nucleon-nucleon collisions are 
neglected in this first approach. Such 
a sudden approximation
is partially relaxed in a more 
realistic situation.}. Thus, the kinematics of the cluster is directly linked
to the kinematics of the nucleons. This defines the {\it 
"total accessible phase-space"} for the cluster $(A_c,Z_c)$ in the medium.
Figure \ref{fig:clus0} displays
the correlation between the position $r_c$ and the kinetic energy $\varepsilon_k$  
of an $\alpha$ particle produced in a Pb target obtained with the
random sampling assumption. We would like to stress that the random assumption
is a generalization of the pioneering work of Goldhaber \cite{Gol74}.  
 
\item {\bf Cluster emission:} we dissociate here the total accessible 
phase-space from the explored 
phase-space because the latter must take into account 
the constraints of the reaction. Indeed, while the first rule described above 
leads to a large set of configurations, all configurations will not 
necessarily lead to the emission of a cluster.
Two constraints can be identified. The first one, 
which is independent of the entrance 
channel, is due to the mutual interaction between the cluster and 
the heavy emitter
\footnote{Since, we are 
considering here rather low beam energy leading to small available energy, 
we do not expect that two 
clusters are emitted at the same time, thus the outgoing channels are 
essentially binary. In addition, 
the use of a heavy target is very helpful since in that case, due to 
the small available energy in 
entrance channel, no particle can be emitted in the secondary decay 
stage. Therefore, in experimental data, 
detected clusters are issued from the pre-equilibrium stage only.}. 
Figure \ref{fig:clus2} shows an example of such an interaction 
in the case of an $\alpha$ particle 
and a $Hg$ 
nucleus. In a classical 
picture, the cluster cannot escape from the heavy nucleus if its 
energy is below the emission barrier. Let $V_{A+A_c}(r_c)$ be 
the interaction potential and $V_B$ 
the associated barrier. We have the 
"local" condition\footnote{Due to the large mass assymetry, 
it is assumed for simplicity that the heavy target is at rest in the 
laboratory frame.}
\begin{eqnarray}
\varepsilon_k(r_c) \geq V_B - V_{A+A_c}(r_c) 
\end{eqnarray}    
leading to a lower limit on the cluster kinetic energy. \\
The second constraint is directly dependent on the reaction type and is 
due to the energy 
balance. Indeed, the accessible configuration is further reduced due to 
the total energy available in the reaction. 
In the simplified scenario presented here (accounting for the fact 
that the initial nucleon is absorbed),
we have the following inequality 
\begin{eqnarray}
E_{B} - Q - V_{A+A_c}(r_c) \geq \varepsilon_k(r_c)
\end{eqnarray}       
which gives an upper limit. Here $E_B$ denotes 
the incident energy while $Q$ is the 
Q-value of the reaction. It is worth to notice 
that the second condition depends not only 
on the beam energy but also on the configuration itself.
Therefore, only a fraction of the total phase-space accessible 
for the cluster will indeed lead to emission in the continuum. 
This fraction corresponds to the {\it "explored phase-space"} 
which takes into
account the energy constraints induced by the reaction. 
\end{itemize}      
These two constraints are shown in Fig. \ref{fig:clus2} (top left)
for a proton-induced reaction at $E_B = 39$ MeV. 
There, an $\alpha$ particle can only be emitted in a small 
interval of kinetic energy (called "escape window"
in the following) leading to a significant 
restriction in phase-space. The fraction of the phase-space available for the
cluster emission is 
shown in Figure 
\ref{fig:clus2} (top-right). According to the 
energy constraint, all configurations 
between the two lines lead to the emission of an $\alpha$ particle. 

\subsection{Direct application of the rules}

\begin{figure}[tbph]
\includegraphics[height=13.cm,angle=-90.]{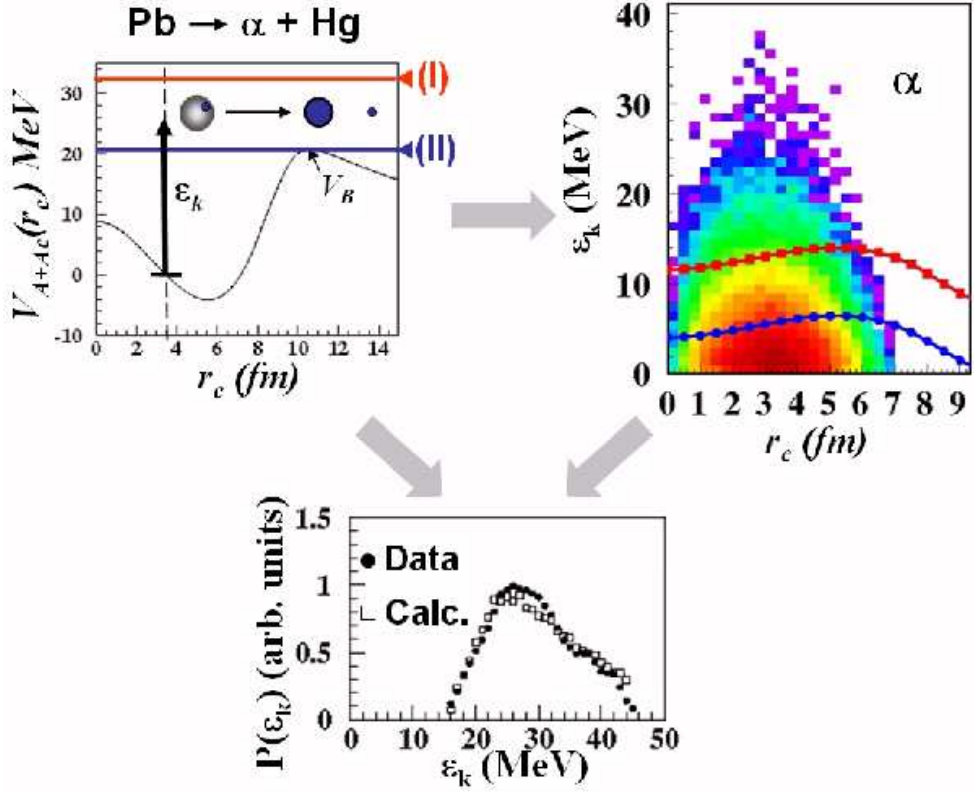}
\caption{ Top-left: Two-body Potential between the $\alpha$ and the 
emitter. Above the line (I), the 
cluster cannot be emitted. This upper limit is directly given by 
the energy balance of the reaction.
Below the line (II), the cluster cannot overcome the barrier (since here, 
quantum tunnelling is not considered). In between the two lines, there is 
a small 
"escape window" for the emission of the cluster.
Top-right: Total available phase-space of the cluster. This latter  
is significantly 
reduced due to the energy constraint. The two curves correspond 
respectively to the lower and 
upper limit in the kinetic energy. 
Bottom: Calculated kinetic energy distribution (open squares) 
of the $\alpha$ particle obtained by propagating each 
configuration in the "escape window" up to infinity. The 
calculated spectrum is compared 
with the experimental data (black circles).
}
\label{fig:clus2}
\end{figure}

\begin{figure}[tbph]
\includegraphics[height=10.cm]{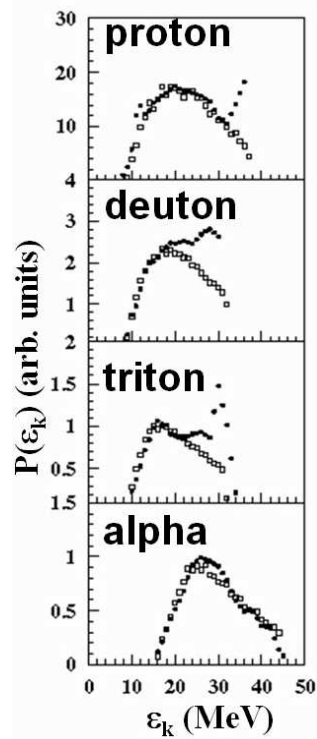}
\caption{ From top to bottom, calculated kinetic energy distributions 
(open squares) obtained
for proton, deuteron, triton and alpha particles. Distributions are  
compared to experimental 
data (open circles).}
\label{fig:clus4}
\end{figure}

The prescription described above are  
compatible with experimental 
data as shown in the case of a proton-induced reaction at $E_B=39$ MeV.
Assuming that the proton is absorbed by the target, a 
Monte-Carlo sampling (using
the cluster creation rules) of the $\alpha$ particle is obtained 
in order to obtain 
the initial configurations in the 
medium. Then, using the emission rules, only those configurations 
allowed by the energy constraint are conserved. 
Last, each conserved configuration is propagated in the target 
potential. Thus, 
the $\alpha$ kinetic energy distribution is obtained and successfully 
compared with the experimental data (from \cite{Ber73}). This is shown 
in Figure \ref{fig:clus2} (bottom
part) where the calculated spectrum 
(open square) is compared to experimental data (filled circles). 
A similar agreement is found for the emission of protons, deuterons 
and tritons (see Fig. \ref{fig:clus4}).
However in this case, direct 
reactions are also present in the 
experimental data leading to an additional contribution at high energy.        

\section{Towards nuclear reactions}

Direct application of the rules to the simplified 
three steps scenario described above 
allows only qualitative comparisons with data. 
In order to provide quantitative comparisons,  
additional effects must be considered. Two 
phenomenological 
models (called n-IPSE\footnote{n-IPSE: nucleon-Ion Phase-Space Exploration} 
\cite{Lac05}  
and HIPSE\footnote{HIPSE: Heavy-Ion 
Phase-Space Exploration} \cite{Lac04} 
based on the very same assumptions have been developed and confronted 
with the experimental data. 
We only give here the physical effects that have been added on top of the rules: 
\begin{itemize}
\item {\bf In medium nucleon-nucleon collisions:} At energies below the Fermi energy, 
the effect of in medium two-body collisions is small. However, as 
the beam energy increases, such collisions must be taken 
into account. Accordingly, the initial Thomas-Fermi 
distributions 
are distorted by two-body effects.
\item {\bf Influence of the impact parameter:} In nucleus-nucleus collisions, 
geometrical aspects associated 
with the impact parameter
are accounted for by using a participant-spectator picture. 
In nucleon-induced reactions, this picture 
is replaced by the "influence area" (equivalent to the participant region) 
notion which defines the number of nucleons of the target affected by 
the projectile. 
\item {\bf One-body dissipation and nucleon absorption:} Depending on 
the incident energy, particles 
are exchanged by the two partners of the reaction: This is treated by means of a
phenomenological parameter (see \cite{Lac04}). In nucleon 
induced collisions, this 
process is replaced by a probability 
that the incoming nucleon be absorbed by the target.  
\item {\bf Application of {\it cluster formation rules} and 
"nucleosynthesis" in the medium:} We have described 
previously a rule to obtain cluster properties when a single cluster 
is formed in the medium. 
However, the number of clusters
is not a priori fixed. In order to solve this difficulty in both
models, a coalescence algorithm is used 
to form clusters starting from the nucleons in the participant region. In
the 
nucleon-induced reaction case, it was 
possible to show explicitly that this coalescence is equivalent 
to a random sampling assumption.     
\item {\bf Application of {\it cluster emission rules} and Final-State
 interaction (FSI)}: After the 
coalescence stage, many configurations are accessible. However
due to the energy-balance generalized 
to the many cluster case, only part of the accessible phase-space 
is really explored. In addition,  
if the relative energy between two clusters is lower than the barrier 
associated with their mutual interaction
they will not separate during the expansion. In a realistic 
model, the recombination 
of fragments is allowed. In HIPSE, possible "re-fusion" of fragments 
is accounted 
for before the freeze-out 
configuration is reached. This process can lead to important FSI's and may 
relax completely the 
participant-spectator picture. For instance, the quasi-target and 
the quasi-projectile can fuse.
\item {\bf Freeze-out and the after-burner stage:} When the available energy 
is large, fragments are excited 
and once the chemical and thermal freeze-out are reached, the possible 
in-flight 
de-excitation of each cluster must be taken into account. This  
induces a complex mixing of 
pre- and post-equilibrium emission. 
\end{itemize}
More details on technical aspects can be found in ref. \cite{Lac04,Lac05}. 
The important point we would like to 
stress is that rather different experimental data can be described using the 
same hypothesis on the production and the emission of clusters. 

\section{Contact with experimental data:  two illustrative examples}

\begin{figure}[tbph]
%\begin{center}
\includegraphics[height=17.cm,angle=-90.]{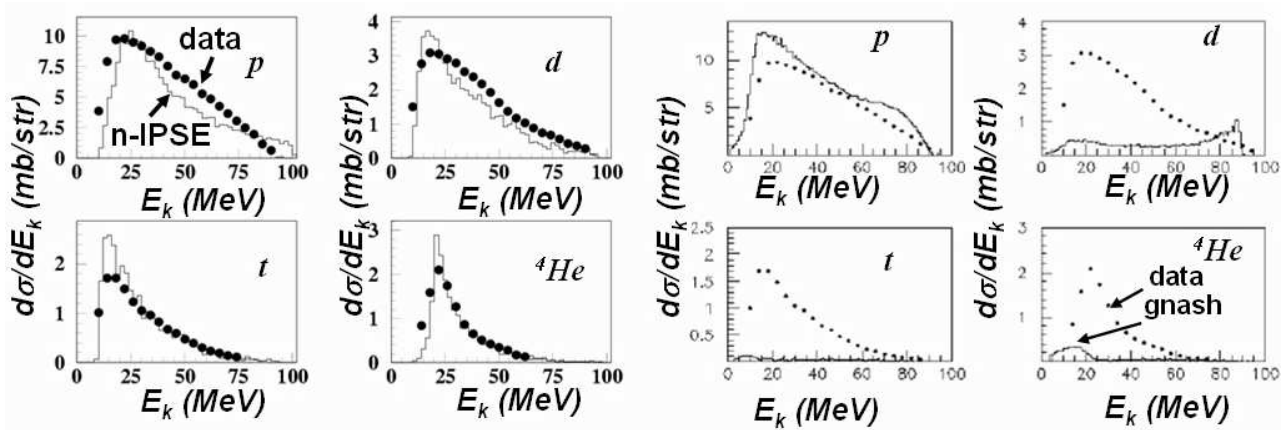}
%\end{center}
\caption{ Left: Kinetic energy differential cross-section of 
proton, deuteron, triton
and alpha particles, obtained with the n-IPSE model
calculation (solid line) for neutron induced reaction on $^{208}$Pb at beam
$E_{B}=96$ MeV (from {\protect\cite{Bli04}}). Right:  
distributions obtained using the GNASH model{\protect \cite{You92}}.}
\label{fig:clus5}
\end{figure}

A detailed comparison of the models with the experimental data 
can be found in \cite{Lac04,Lac05}. Here, we first concentrate on
nucleon-induced reactions. Figure \ref{fig:clus5} shows a comparison 
between the  
kinetic energy differential cross-section of light clusters 
calculated with n-IPSE and data \cite{Bli04}. 
Note that there is no normalization 
between the data and
the calculation. As 
a reference, we also show (right part of the figure) 
the calculated spectra obtained with GNASH\cite{You92}. A good agreement
between
the results of n-IPSE and the experimental data 
is obtained. This is true for 
a wide range of beam energy
from $37$ MeV to $135$ MeV in both proton and neutron induced reactions 
on medium and heavy nuclei.      

\begin{figure}[tbph]
%\begin{center}
\includegraphics[height=10.cm,angle=-90]{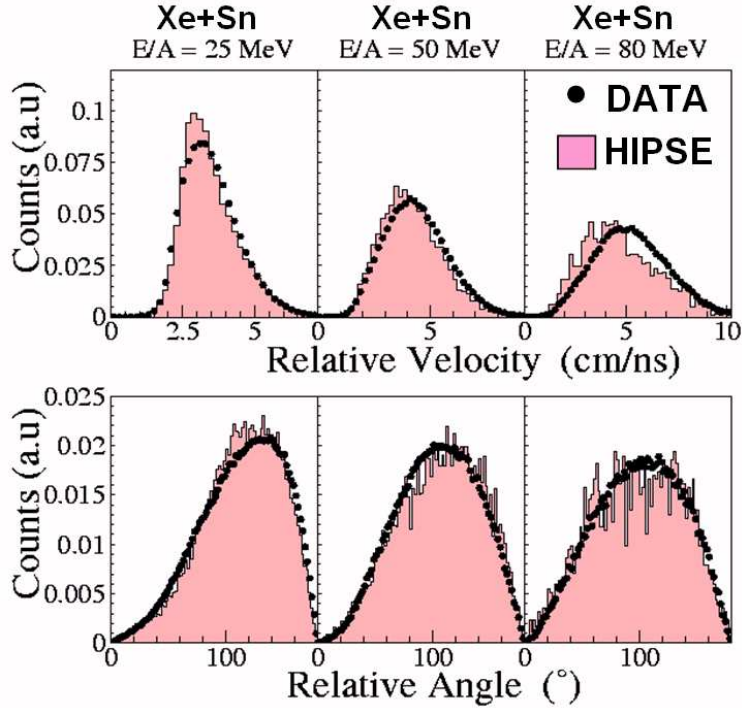}
%\end{center}
\caption{ Distributions of the relative velocity (top) and the relative
 angle (bottom) between the three largest 
fragments taken two-by-two for the $Xe+Sn$ system at three 
different beam energies. From left to right, the beam 
energies $E_B=25$, $50$ and $80$ MeV/A are considered. 
In each case, the calculated spectra are compared with 
the INDRA data (filled circle). Events 
considered here correspond to well detected 
events (80 $\%$ of the total charge and impulsion). The same 
selection is used for the "filtered" HIPSE events. }
\label{fig:clus6}
\end{figure}
Concerning nucleus nucleus reactions, a systematic comparison with the 
INDRA data\cite{Hud03} have demonstrated 
that the HIPSE model is able to reproduce not only the average
 properties \cite{Lac05} but also the fluctuations of the 
experimental observables\cite{Van03}. It appears that 
besides mean properties 
and fluctuations, "internal" correlations inside each event
are also correctly reproduced as shown in
Figure \ref{fig:clus6} \cite{Van03-2} where the distribution 
of the relative velocity (top) and the 
relative angle (bottom) between the three largest 
fragments taken two-by-two  
are presented for the reaction Xe+Sn at three 
different beam energies ($E_B=25$, $50$, $80$ MeV/A). In 
each case,
the calculated spectra are compared 
with the INDRA data \cite{Hud03}. The 
very good agreement between 
HIPSE and the INDRA data gives additional proof of the 
compatibility 
between the rules for the production and the emission of complex particles 
and the experimental observables. Note that a similar 
agreement has been is found in 
different symmetric systems\footnote{Please note that, in both 
models, a few free parameters are used: they are related
to the description of the participant region, the nucleon-nucleon 
collision rate , 
the exchange and absorption nucleon processes. 
Such parameters depend only on 
the beam energy. This means, that a single 
set of parameters is used to reproduce simultaneously 
nucleon reactions on Fe, Pb and U. Similarly, 
the parameters adjusted 
on Xe+Sn reactions have been used to the Ni+Ni and the Au+Au cases giving
reasonable agreement with the data.}.

\section{Conclusion}
We have described simple rules
 that may be used to describe the pre-equilibrium emission 
of clusters in the course of nuclear reactions. These rules are based on 
a random sampling of the nucleons taking into account the Fermi motion 
and a proper account of nuclear effects 
as well as the conservation laws. Using these
rules leads to a good agreement with data obtained from nucleus-nucleus reactions 
around the Fermi energy and surprisingly also for nucleon-nucleus
reactions. 

We would like to mention, that even if the complete randomness 
hypothesis appears compatible 
with the experimental data, this do not give any indication on the physical 
origin of randomness and several 
effects can be invoked: phase-transition, turbulence, 
self-organized criticality, quantum decoherence ... 

{\bf Acknowledgments}
We thank warmly the INDRA collaboration for permission to 
use its  data. We would like to thank V. Blideanu, O. Lopez, A. Van Lauwe and 
E. Vient for their collaboration in this work.

\end{document}